\documentclass[amssymb,aps, prl,superscriptaddress, 10pt]{revtex4-2}

\usepackage{amsmath}
\usepackage{amssymb}
\usepackage{amsfonts}
\usepackage{color}
\usepackage[pdftex]{graphicx}
\usepackage{epstopdf}
\usepackage{soul}
\usepackage{xcolor}
\usepackage{multirow} 
\usepackage{cases}

\newcommand{\Qjt}{Q_j^{\tau}}

\newcommand{\Qje}{Q_j^{(e)}}
\newcommand{\Qjm}{Q_j^{(m)}}
\newcommand{\Qjtn}{\Qjt(n)}
\newcommand{\Qjtk}{\Qjt(k)}
\newcommand{\xrt}{x^{\tau }_j}

\newcommand{\xrm}{x^{(m)}_j}

\newcommand{\Bjt}{B_j^{\tau}} 
\newcommand{\Qz}{Q_{j,0}^{\tau}} 
\newcommand{\kjts}{k_{j,s}^{\tau}}
\newcommand{\xrd}{x_1^{(m)}}

\begin{document}

\title{Quality factor of dielectric optical resonators} 

\author{Xavier Zambrana-Puyalto}
\email{xavislow@protonmail.ch}
\affiliation{Department of Physics, Technical University of Denmark, Fysikvej, DK-2800 Kongens Lyngby, Denmark}

\author{Søren Raza}
\email{sraz@dtu.dk}
\affiliation{Department of Physics, Technical University of Denmark, Fysikvej, DK-2800 Kongens Lyngby, Denmark}

\begin{abstract}
We show analytically that the quality ($Q$) factor of magnetic and electric Mie modes in a lossless dielectric spherical resonator with high refractive index ($n \gg 1$) scales as $n^{2j+1}$ and $n^{2j+3}$ respectively, where $j$ denotes the multipolar order. We numerically validate these results and show that our high-$n$ analytical relation is accurate for the dipolar modes when $n>5$. For higher multipolar orders, the analytical relation becomes valid for increasingly lower $n$. We study the dependence of the $Q$ factor on absorption losses and determine a general functional form that describes the $Q$ factor for all Mie modes for any complex refractive index. Finally, we observe that this functional form predicts a multipolar-dependent singular value of optical gain which gives rise to a lasing condition with an infinite $Q$ factor. 
\end{abstract}

\maketitle

\section{Introduction}
The investigation of optical resonators, characterized by their ability to confine and enhance electromagnetic fields at the nanoscale, is a fundamental pursuit in the realm of photonics~\cite{Saleh1991}. Understanding the resonant behaviour of these structures is essential for designing devices with enhanced performance~\cite{Jahani2016}. In this regard, the quality ($Q$) factor is one of the most important characteristics as it provides a measure for the light recirculation in an optical resonator. The $Q$ factor of a resonance is typically defined as $Q = \nu_0/\Delta \nu$, where $\nu_0$ is the central resonant frequency, and $\Delta \nu$ is the full width at half maximum (FWHM) of the resonance~\cite{Saleh1991}. A high $Q$ factor indicates that the resonance is sharp and as a result, the trapped light in the resonator barely leaks out~\cite{Oraevsky2002}. 
High $Q$-factor optical resonators based on dielectric materials are seen as the building blocks for the next generation of optical circuitry at the nanoscale~\cite{Barton2020}. These resonators have become instrumental in a range of applications, such as enhancing nonlinear optical processes~\cite{Koshelev2020}, enabling molecular sensing~\cite{Tittl2018}, generation of non-classical light~\cite{Santiago-Cruz2022}, and facilitating strong light-matter coupling~\cite{Spillane2005,Faucheaux2014}.

Despite its importance, there is no general relation for the $Q$ factor even for the most fundamental dielectric Mie resonator; the sphere. Although research has shown certain bounds on the $Q$ factor~\cite{Schuller2009} as well as established general relations for the scattering properties of dielectric resonators~\cite{Hulst1957}, $Q$ factor relations have been numerically identified for only a few specific Mie modes~\cite{Baranov2017,Rybin2017}. This is in contrast to the case of metallic sphere resonators, where a general relation for the $Q$ factor has been established~\cite{Wang2006}, largely due to the applicability of the quasistatic approximation, which simplifies the analysis considerably. This approach, however, is not suitable for dielectric resonators, requiring new methods to address the complexities introduced by Mie theory for dielectric materials. Establishing a general relation for the $Q$ factor would be an important guide towards the design of dielectric resonators with optimized performance.


In this Letter, we determine a general relation for the $Q$ factor for any resonant multipolar mode of a spherical dielectric Mie resonator. We analytically demonstrate that the $Q$ factor of a dielectric resonator with a high refractive index $n$ scales as $n^{2\left( j+1  \right)+\tau}$, where $j$ and $\tau$ denote the multipolar order and parity sign of the resonance, i.e., $ \tau = -1 (+1)$ for magnetic (electric) modes~\cite{Tung1985}. We numerically verify our analytical result, and observe that the relation is applicable down to $n \approx 5$ for the dipolar modes and for even lower $n$ as the multipolar order of the resonance increases. The strong power-law dependence on the refractive index has important consequences for the $Q$ factor, especially for higher-order modes, where the $Q$ factor can reach $\sim10^2$ with refractive index values available with current dielectric materials covering the visible spectral range~\cite{Baranov2017,Khurgin2022,Svendsen2022}. We also numerically study how the $Q$ factor of a general multipolar mode depends on the absorption losses, quantified by the extinction coefficient $k$, and identify a general trend when we use $ k \cdot n ^{2(j+1/2)+\tau}$ as independent variable. This analysis reveals a general expression for the $Q$ factor of any Mie mode for any complex refractive index $n+ik$. Finally, we observe that our result can be extended to negative values of $k$, i.e., to materials with optical gain. Our formulation shows that each Mie mode has a unique singular gain condition where the $Q$ factor becomes infinite, and that this condition is determined by a single numerical parameter. We find that the singular gain condition corresponds to the threshold for single-mode lasing~\cite{Mostafazadeh2011,Tiguntseva2020,Farhi2024} and completely determines how a given mode is affected by absorption losses. 

\section{Analytical demonstration of $ \mathbf{\Qjtn} $ for a lossless material} \label{sec:analytic}
The scattering of light from a dielectric sphere with radius $R$ and complex refractive index $n+ik$ surrounded by a lossless medium is exactly described by Mie theory~\cite{Bohren1983}. In Mie theory, the electromagnetic fields are expanded in spherical multipoles that describe the eigenmodes of the sphere. The scattering properties of the sphere are given by the so-called Mie coefficients, which determine the eigenfrequencies of the system. Each Mie coefficient has a numerable infinite number of eigenfrequencies~\cite{Zambrana2015,Oraevsky2002}. In this work, we consider only the lowest energy eigenfrequency associated to each multipolar mode, as this is the one with the largest $Q$ factor~\cite{Oraevsky2002} and is the resonance typically observed in experiments involving high-refractive-index nanostructures. 

We now outline the analytical derivation of the functional form of the $Q$ factor, $\Qjtn$, for both parities $\tau = \pm 1$ and any multipolar order $j$ for a lossless dielectric sphere ($k=0$). We consider first the $Q$ factor associated to the magnetic modes ($Q_j^{(m)} \equiv Q_j^{\tau=-1}$), which are described by the magnetic Mie coefficients $b_j(n,x)$ that depend on the real part of the refractive index $n$ and the size parameter $x=2\pi R/ \lambda$. The magnetic Mie coefficients are rewritten in terms of the magnetic Mie phase angle $\beta_j(n,x)$ as~\cite{Hulst1957}
\begin{equation}
    b_j(n,x) = \frac 12 \left(1 - e^{2i\beta_j(n,x)} \right),
\end{equation}
where the Mie phase angle is given by
\begin{equation}
\tan \beta_j(n,x) = - \dfrac{n \psi'_j(y)\psi_j(x)-\psi_j(y)\psi'_j(x)}{n \psi'_j(y)\chi_j(x)-\psi_j(y)\chi'_j(x)}.
\label{eq:tanBeta}
\end{equation}
Here, $\psi_j$ and $\chi_j$ denote the Riccatti--Bessel function of order $j$, and $y=nx$. The prime denotes differentiation with respect to the argument. The benefit of introducing the Mie phase angles is that these functions, in contrast to the Mie coefficients, are by definition real-valued functions for lossless dielectrics. 

In the limit of high refractive indices ($n \gg 1$), the resonance of multipolar order $j$ is given by the resonant size parameter $\xrm = r_{j-1}/n$, where $r_j$ denotes the first zero of the Riccati--Bessel function, e.g., $r_0=\pi$~\cite{Hulst1957}. Since the size parameter is proportional to the frequency, we can compute the $Q$ factor as $\Qjm = \xrm/\mathrm{FWHM}_{\xrm}$. Our aim now is to find the functional form of the full-width at half-maximum $\mathrm{FWHM}_{\xrm}$. To this end, it is worth noting that the real part of the magnetic Mie coefficients, which is directly linked to the scattering and extinction cross sections, can be expressed as $\mathrm{Re}(b_j)=\dfrac{1}{2} \left[ 1- \cos(2\beta_j) \right]$. On resonance, the Mie coefficient takes the value $\mathrm{Re}(b_j) = 1$, which implies that the phase angle is $\beta_j=\frac{\pi}{2}$.
\begin{figure}
	\centering
	\includegraphics[width=0.8\columnwidth]{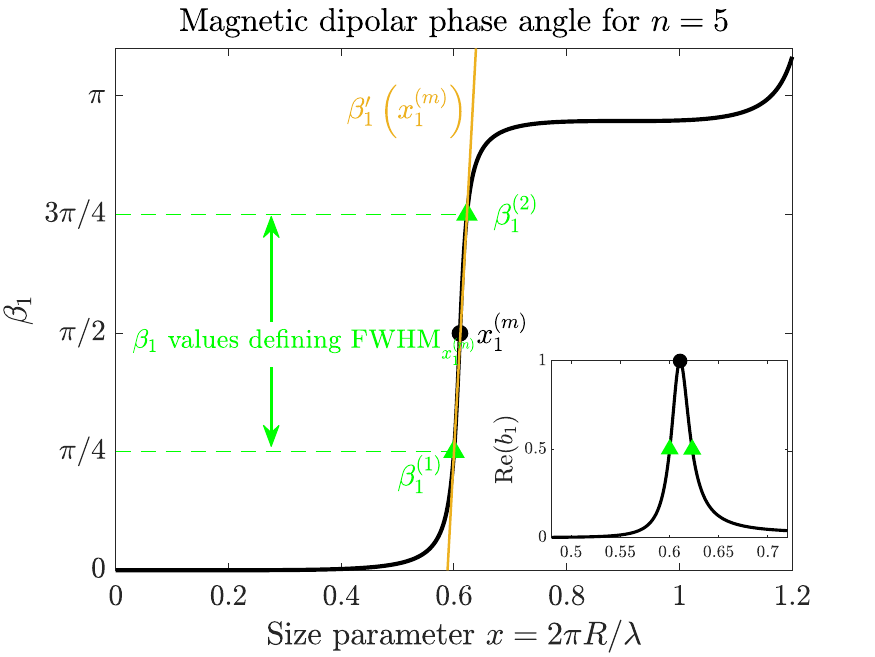}
	\caption{Dipolar magnetic Mie phase angle $\beta_1$ as a function of the size parameter $x$ for a dielectric sphere with refractive index $n=5$. At the resonant condition, $\beta_1(x_1^{(m)})=\pi/2$, which corresponds to the resonant condition $\mathrm{Re}\left( b_1 \right)=1$ (see inset plot). The FWHM of the resonance is determined by two size parameters that make $\beta_1^{(2)}=3\pi/4$ and $\beta_1^{(1)}=\pi/4$ (green triangles). The derivative of the Mie phase angle on resonance, $\beta'_1(x_1^{(m)})$, is shown in orange. \label{fig:beta1}}		
\end{figure}
%
The FWHM of the resonance is then computed as the difference between the two size parameters that make $\mathrm{Re}(b_j)=1/2$, which gives the phase angles $\pi/4$ and $3\pi/4$. In the high-$n$ limit, the Mie phase angle is close to linear around the resonance, allowing us to relate the FWHM with the derivative of the phase angle as $\beta'_j\left(\xrm\right)=\frac{\pi/2 }{\mathrm{FWHM}_j}$, thus yielding the following relation for the $Q$ factor
\begin{equation}
Q^{(m)}_j= \dfrac{2}{\pi} \xrm \beta'_j \left( \xrm  \right).
\label{eq:Qbeta}
\end{equation}

All of these features are graphically shown in Fig.~\ref{fig:beta1}, where the magnetic dipolar Mie phase $\beta_1$ is plotted as a function of the size parameter for a particle of $n=5$. We have highlighted the resonant condition $\xrd$ with a black dot, and the two $\beta_1$ values that define the FWHM with green triangles. Note that the FWHM is the distance between the size parameters that yield $\beta_{1}^{(2)}$ and $\beta_1^{(1)}$. Furthermore, we also display in orange the derivative of the phase angle $\beta_1$ at the resonant condition. An inset with a plot of $\mathrm{Re}\left( b_1 \right)$ is also included that links the Mie coefficient to the Mie phase angle.

The derivation of the functional form of the derivative of the phase angle on resonance, i.e., $\beta'_j \left( \xrm  \right)$, is presented in detail in the Supplementary Material (SM). Here, we simply state the result which is
\begin{equation}
    \beta'_j(\xrm)  = \frac{(2j-1)!!(2j+1)!!}{2j+1} r_{j-1}^{-2j} n^{2j+2}, \label{eq:betaprime}
\end{equation}
where $!!$ denotes the double factorial. Inserting Eq.~(\ref{eq:betaprime}) into Eq.~(\ref{eq:Qbeta}), we finally obtain
%
\begin{equation}
\begin{aligned}
\Qjm & = K_j^{(m)} n^{2j+1}, \label{eq:Qjm} \\
K_j^{(m)} & = \dfrac{2}{\pi} \dfrac{(2j-1)!!(2j+1)!!}{2j+1} r_{j-1}^{-2j+1}.
\end{aligned}
\end{equation}

Similar calculations can be carried out for the electric Mie coefficient $a_j$, which can be rewritten by means of the electric Mie phase angle $\alpha_j$. Proceeding along the lines (see the SM for details), we find that the $Q$ factor of the electric modes are given by
\begin{equation}
\begin{aligned}
\Qje & = K_j^{(e)} n^{2j+3}, \\
K_j^{(e)} &= \dfrac{2}{\pi} \dfrac{(2j-1)!!(2j+1)!! j^2}{2j+1} r_{j}^{-2j-1}. \label{eq:Kje}
\end{aligned}
\end{equation}

The $Q$ factors of the electric and magnetic modes, given by Eqs.~(\ref{eq:Qjm}-\ref{eq:Kje}), can be generalized by introducing the parity sign parameter $\tau$ as:
\begin{equation}
\begin{aligned}
\Qjt & =  K_j^{\tau} n^{2\left( j+1  \right)+\tau}, \label{eq:Qjt} \\
 K_j^{\tau} & = \dfrac{2}{\pi} \dfrac{(2j-1)!!(2j+1)!! } {2j+1} \left( \dfrac{j \left( \tau +1  \right)}{2} \right)^2  r_{(2j+\tau - 1)/2}^{-2j-\tau}, 
\end{aligned}
\end{equation}
where $\tau=-1$ ($\tau=+1$) for magnetic (electric) modes. Equation (\ref{eq:Qjt}) is one of the main results of this paper and shows that the $Q$ factor of any multipolar Mie mode in a lossless spherical resonator with high refractive index depends solely on the refractive index $n$ of the resonator. In Table \ref{tab:Qvalues}, the expressions of $\Qjt(n)$ for the first three multipolar orders and both parities are given.
\begin{table}
\begin{center}
\begin{tabular}{ |c|c|c| } 
 \hline
 Multipolar order $j$ & $\Qjm$ & $\Qje$   \\ 
 \hline
 $j=1$ (Dipole) & $Q_1^{(m)} \approx 0.20 n^3$  & $Q_1^{(e)} \approx 0.0070 n^5$ \\ 
  $j=2$ (Quadrupole) & $Q_2^{(m)} \approx 0.063 n^5$ & $Q_2^{(e)} \approx 0.0036 n^7$ \\ 
 $j=3$ (Octupole) & $Q_3^{(m)} \approx 0.022 n^7$  & $Q_3^{(e)} \approx 0.0016 n^9$ \\
 \hline
\end{tabular}
 \caption{Numerical values of $\Qjm$ and $\Qje$ for the dipole, quadrupole, and octupole modes ($j=1,2,3$). \label{tab:Qvalues}}
\end{center}
\end{table}

\section{Numerical calculation of $ \mathbf{\Qjtn} $ for a lossless material} \label{sec:numQn}

To verify our analytical relations, we simulate the exact behaviour of $\Qjtn$ for two magnetic and electric Mie modes in the range of $n \in \left[2, 20 \right]$. We obtain the $Q$ factor associated with a certain Mie mode (defined by $j,\tau$) for each refractive index by determining the first (lowest energy) pole of the Mie coefficient $\xrt$. The pole is a complex number, so lowest energy refers to lowest $\mathrm{Re} ( \xrt)$. We compute the $Q$ factor as $\Qjt=\frac{\mathrm{Re} ( \xrt)}{2 \mathrm{Im} ( \xrt)}$ \cite{Oraevsky2002}. This process is repeated until the whole refractive index range is covered (see details in SM). In the SM we also show the results for $Q$ factor in the refractive index regime $n \in \left(1, 2 \right]$. 
\begin{figure}
	\centering
	\includegraphics[width=0.8\columnwidth]{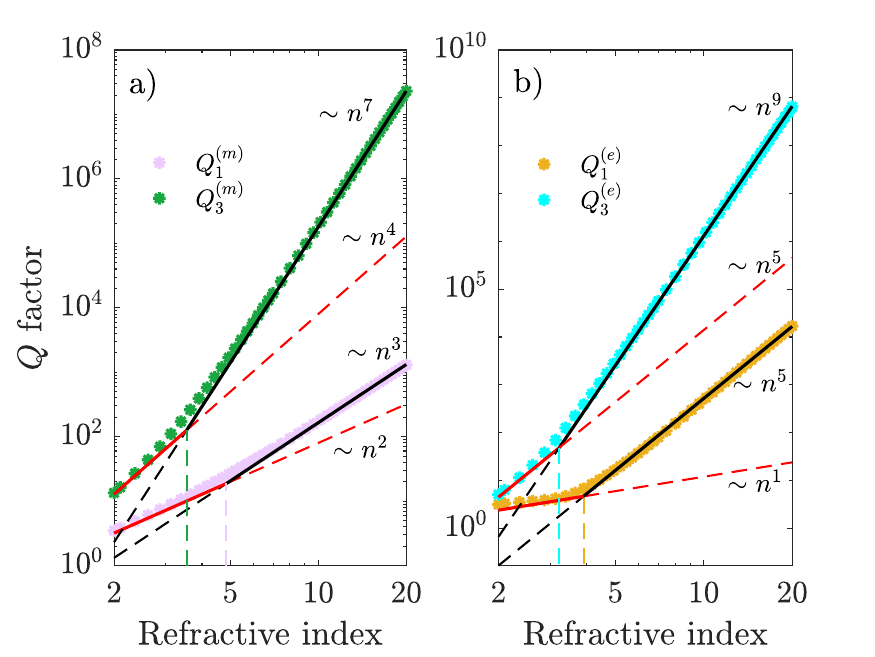}
	\caption{$Q$ factor as a function of the refractive index. a) Log-log plot of the $Q$ factor of the magnetic dipolar $j=1$ (pink) and octupolar $j=3$ modes (green). b) Log-log plot of the $Q$ factor of the electric dipolar $j=1$ (orange) and octupolar $j=3$ modes (cyan). In both (a-b), the refractive index is varied from $2$ to $20$. The high-$n$ behaviour scaling $ \sim n^{2\left( j+1  \right)+\tau}$ is plotted in black, and the low-$n$ scaling is plotted in red. The scaling is plotted as a straight (dashed) line for the points where the corresponding scaling (does not) holds. The dashed vertical lines  reference the values of $n$ where a transition between the high-$n$ and low-$n$ scaling occurs.
	  \label{fig:Qnreal}}		
\end{figure}

Figure~\ref{fig:Qnreal}a displays a log-log plot of the simulated $Q$ factors for the dipole ($j=1$) and octupole modes ($j=3$). We observe that the results of our numerical simulations perfectly follow the theoretical relations given by Eq.~(\ref{eq:Qjt}) for values of $n > 5$ (black lines). We also observe that as the refractive index decreases, the scaling of $\Qjm$ starts deviating from the analytical high-$n$ limit and instead adopts a weaker dependence on the refractive index (red lines in Fig. \ref{fig:Qnreal}a). This weaker dependence can be modelled as $\sim n^c$ with an exponent $ c  < 2j+1  $. The observations for the $Q$ factor of the electric modes are analogous (see Fig. \ref{fig:Qnreal}b). As a result, the scaling of $\Qje$ for low refractive indices is $\sim n^c$ with a modified exponent $ c  <  2j+3  $ . As it was done in Eq.~(\ref{eq:Qjt}), we can generalize the behaviour of $\Qjm$ and $\Qje$ as $\Qjtn = K_j^{\tau} n^{2\left( j+1  \right)+\tau}$ for $n > 5$, and $\sim n^c$ with $ c  < \left[  2\left( j+1  \right)+\tau \right] $ for $n$ values close to $n = 2$. Since Fig. \ref{fig:Qnreal} is a log-log plot, it means that the $Q$ factor approximately behaves as the sum of refractive-index power-law terms, i.e., $  \Qjtn  \sim  K_j^{\tau} n^{2\left( j+1  \right)+\tau} + C_j^{\tau} n^{c} $. This is a consequence of the fact that the logarithm of a sum is approximately given by
\begin{equation}
\log \left( x + y \right) \approx  \begin{cases} \log(x), & x \geq y \\ \log(y), & y \geq x \end{cases}
\label{eq:logSum}
\end{equation} 

Although materials exhibiting refractive indices above $5$ in the visible are lacking, such materials are readily available at longer wavelengths, such as in the infrared \cite{Baranov2017, Khurgin2022, Svendsen2022, Doiron2022}. The analytical high-$n$ relation is accurate at increasingly lower refractive index ($n$) as the multipolar order of the mode ($j$) is increased. In fact, for octupolar modes, we observe that the high-$n$ relation is accurate for refractive index values that are available for operation in the visible spectral range. This is indicated by the vertical dashed lines in Fig. \ref{fig:Qnreal}(a-b), which represent threshold refractive index values for which a lower refractive index scaling of the $Q$ factor $\sim n^c$ is more appropriate than the high-$n$ limit. It is worth noting that the strong refractive-index dependence of higher-order multipoles means that $Q$ factors of $\sim10^2$ can be achieved through both the electric and magnetic octupolar modes with a modest refractive index of 3.5. The fact that such high $Q$ factors are available in the archetypical spherical resonator complements the ongoing efforts on resonator designs based on bound states in the continuum~\cite{Rybin2017}. Finally, Fig.~\ref{fig:Qnreal} also shows that the $Q$ factor of electric modes is greater than that of magnetic modes in the high-$n$ limit (for the same $j$). However, for lower refractive indices, it is just the opposite, i.e., the magnetic modes have higher $Q$ factor. 

\section{The effect of losses}
\begin{figure}
	\centering
	\includegraphics[width=\columnwidth]{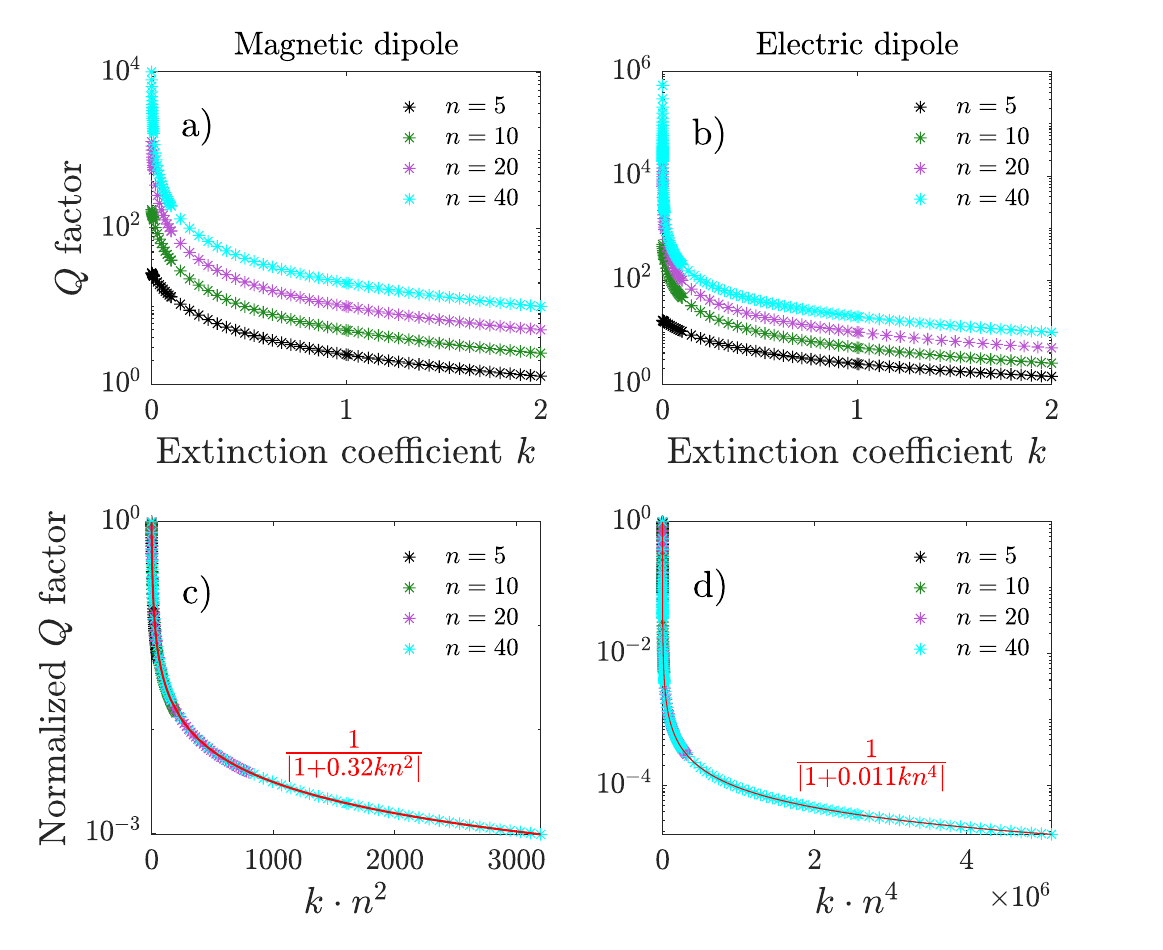}
	\caption{a) $Q$ factor of the magnetic dipolar mode as a function of the extinction coefficient for four different refractive indices, in semilog space. b) Same plots as in a) for the electric dipolar mode. c) Normalized $Q$ factor of the magnetic dipolar mode as a function of $k\cdot n^2$. The normalized $Q$ factor is accurately described by a unique curve (red line) given by Eq.~(\ref{eq:Qlosses}). d) Same plots as in c), but for the electric dipolar mode as a function of $k\cdot n^4$. \label{fig:Qnk}}		
\end{figure}

In this section, we study how losses reduce the $Q$ factor of a Mie mode. We characterize losses via the imaginary part of the complex refractive index, i.e., the extinction coefficient $k$. We compute $\Qjtk$ for both the electric and magnetic dipolar modes, for four different refractive indices and for varying $k$ (see Fig. \ref{fig:Qnk}a-b). See the SM for results on the quadrupolar modes.
We observe that $\Qjt$ is inversely proportional to $k$, i.e., optical losses reduce the quality factor of an optical resonator. We also observe that, in spite of losses, greater values of $n$ still yield larger $Q$ factors. The four curves in Fig. \ref{fig:Qnk}a-b obtained with different refractive indices display a similar decay with increasing $k$, indicating a similar functional form. This suggests that the four curves can be collapsed into a unique curve that describes them all. To determine their functional form, we normalize each curve by its respective lossless $Q$ factor (denoted as $\Qz \equiv \Qjt(n,k=0)$), such that the four curves yield $\Qjt/\Qz=1$ when $k=0$. We find that the four curves collapse into a general curve when $k \cdot n ^{2(j+1/2)+\tau}$ is used as the independent variable, see Fig.~\ref{fig:Qnk}c-d. Our next step is finding a functional form for this behaviour. We observe that $\Qjt \left( k \cdot n ^{2(j+1/2)+\tau} \right)$ has two very clear limits for zero and large values of $k \cdot n ^{2(j+1/2)+\tau}$, and a smooth transition in between. This is similar to the behaviour of the logarithm of a sum, given by Eq.~(\ref{eq:logSum}). We know that $\lim_{k \to 0} \Qjt/\Qz = 1 $, so we need to determine the decay for large $k \cdot n ^{2(j+1/2)+\tau}$. After several trials, we find that the following functional form, which only depends on a single numerical parameter $\Bjt$, perfectly describes our numerical data for the normalized $Q$ factor $\widetilde{Q}_{j}^{\tau}$:
\begin{equation}
\widetilde{Q}_{j}^{\tau}(n,k) \equiv \dfrac{\Qjt(n,k)}{Q_{j,0}^{\tau}(n)}=\dfrac{1}{\vert   1 + B_j^{\tau}  k \cdot n^{2(j+1/2)+\tau} \vert}.
\label{eq:Qlosses}
\end{equation}
We compute $B_j^{\tau}$ for both the magnetic and electric dipolar modes, yielding $B_1^{(m)}=0.32$ and $B_1^{(e)}=0.011$. Figure~\ref{fig:Qnk}c displays the functional form of $  \widetilde{Q}_{1}^{(m)}$ with $B_1^{(m)}=0.32$ on top of the normalized $Q$ factors obtained in Fig.~\ref{fig:Qnk}a. Similarly, Figure~\ref{fig:Qnk}d shows the function $\widetilde{Q}_{1}^{(e)}$ on top of the normalized numerical data displayed in Fig.~\ref{fig:Qnk}b. It is observed that the numerical data matches the functional form given by Eq.~(\ref{eq:Qlosses}), thus validating the expression. Equation~(\ref{eq:Qlosses}) is one of the main results of this article and describes the $Q$ factor dependence on both the real and imaginary parts of the refractive index. Given the lossless $Q$ factor of a certain Mie mode $Q_{j,0}^{\tau}(n)$, it allows us to calculate how the $Q$ factor decreases for any extinction coefficient $k$ with just one constant $\Bjt$. 

We combine this result with the our analytical result for the lossless case, given in Eq.~(\ref{eq:Qjt}), to write up a general expression for the $Q$ factor of any Mie mode for any complex refractive index with $n>5$
\begin{equation}
\Qjt(n,k) = \dfrac{K_j^{\tau} n^{2\left( j+1  \right)+\tau} }{\vert   1 + B_j^{\tau}  k \cdot n^{2(j+1/2)+\tau} \vert},
\end{equation}
where the numerical values of $\Bjt$ are given in Table \ref{tab:Bjt} for the first three multipolar orders, and both parities.
\begin{table}
\begin{center}
\begin{tabular}{ |c|c|c| } 
 \hline
 Multipolar order $j$ & $B_j^{(m)}$ & $B_j^{(e)}$   \\ 
 \hline
 $j=1$ (Dipole) & $B_1^{(m)} \approx 0.32 $  & $B_1^{(e)} \approx 0.011 $ \\ 
  $j=2$ (Quadrupole) & $B_2^{(m)} \approx 0.097 $ & $B_2^{(e)} \approx 0.0049 $ \\ 
 $j=3$ (Octupole) & $B_3^{(m)} \approx  0.035$  & $B_3^{(e)} \approx 0.0022$ \\
 \hline
\end{tabular}
 \caption{Numerical values of $B_j^{(m)}$ and $B_j^{(e)}$ for $j=1,2,3$. \label{tab:Bjt}}
\end{center}
\end{table}

\section{The effect of gain}
\begin{figure}
	\centering
	\includegraphics[width=0.8\columnwidth]{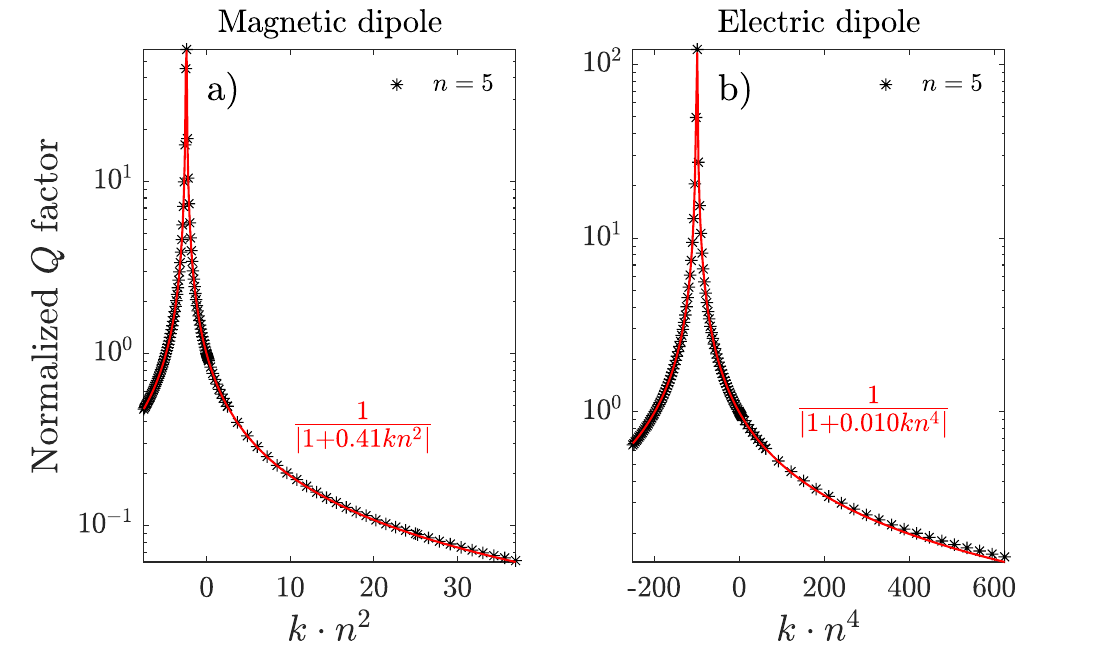}
	\caption{a) Numerical simulations of the normalized $Q$ factor of a magnetic dipolar mode $\widetilde{Q}_{1,N}^{(m)}$ as a function of $k \cdot n ^2$. b) Numerical simulations of the normalized $Q$ factor of an electric dipolar mode $\widetilde{Q}_{1,N}^{(e)}$ as a function of $k \cdot n ^4$. For both plots, the numerical data is plotted as black crosses. The red curves follow the functional form given by Eq.~(\ref{eq:Qlosses}). Both plots are done for a sphere with a refractive index of $n=5$.  \label{fig:Qgain}}	
\end{figure}

Up until now, we have modelled absorption as a positive-valued extinction coefficient $k$. The same formulation can be applied to model the effect of optical gain on the quality factor by using a negative-valued extinction coefficient, i.e., $k<0$. In fact, Eq.~(\ref{eq:Qlosses}) includes an absolute value in the denominator because the $Q$ factor is by definition a positive number. If this was absent, it would have opened the possibility for negative $Q$ factors for certain values of $k$ and $\Bjt$. Inspection of Eq.~(\ref{eq:Qlosses}) reveals that a singular $k$ value can be defined such that $\Qjt$ diverges
\begin{equation}
k_{j,s}^{\tau}=\dfrac{-1}{\Bjt n^{2(j+1/2)+\tau} }.
\label{eq:ks}
\end{equation}
That is, Eq.~(\ref{eq:Qlosses}) predicts that each multipolar mode has an associated singular gain value $\kjts$, given by Eq.~(\ref{eq:ks}), that makes the $Q$ factor diverge. Notice that this singular gain value is a threshold for the system to be in a single-mode lasing condition~\cite{Mostafazadeh2011,Tiguntseva2020,Farhi2024}. To corroborate this finding, we extend our numerical simulations of the $Q$ factor for negative values of $k$. In Fig. \ref{fig:Qgain}, we show the results of these simulations for the electric and magnetic dipolar modes for $n=5$. See the SM to find the same curves for other refractive indices. In Fig. \ref{fig:Qgain}, we also display (in red) the functional form given by Eq.~(\ref{eq:Qlosses}). Again, we observe that Eq.~(\ref{eq:Qlosses}) perfectly describes the numerical simulations of the $Q$ factor. However, the $\Bjt$ parameter has changed slightly for both functions. This indicates that $\Bjt$ has a weak dependence on $n$, and we find that the this dependence becomes negligible for $n>5$ (see SM). Despite this, it is worth noting that Eq.~(\ref{eq:Qlosses}) perfectly captures the physics of the problem, as it has allowed us to predict the unexpected phenomenon of a singular gain value $\kjts$.  

It is interesting to note that there is a bijective relation between $\kjts \Longleftrightarrow \Bjt$. That is, given a certain Mie mode, we can fully determine its singular gain condition by knowing how its lossless $Q$ factor $\Qz$ is affected by losses. Or in other words, the knowledge of the singular gain condition of a Mie resonator $\kjts$ completely determines how the $Q$ factor associated to this Mie mode is affected by losses. This property is reminiscent of the Weierstrass factorization~\cite{Grigoriev2013,Remi2017}, which states that knowledge of the zeros and poles of a complex function is enough to determine the function in the complex plane. That is, the poles of $\Qjt$ in the $(n,k)$ plane, which are given by Eq.~(\ref{eq:ks}), determine its functional form. It is worth noting that $\Qjt$ is symmetric around the singular point. However, there is a subtle difference between the two decaying branches of $\Qjt$, one from the singular gain value to $\infty$, and the other from the singular gain value to $-\infty$. The difference is that the imaginary part of the poles retrieved to build the absorption branch (from $\kjts$ to $\infty$) has a negative value, whereas the poles used to build the gain branch have a positive imaginary part. The two branches are separated by the singularity, which has a zero imaginary part, thus yielding an infinite $Q$ factor.

\section{Conclusion}
In conclusion, we have analytically proven that the $Q$ factor of a Mie mode of a lossless spherical particle scales as $\Qjt(n)  =  K_j^{\tau} n^{2\left( j+1  \right)+\tau} $ for large values of the refractive index $n$, with $K_j^{\tau} $ being a constant that depends on the multipolar order $j$ and the parity $\tau$ of the mode. We have numerically calculated $\Qjt(n)$ for different Mie modes. We have corroborated that the magnetic (electric) mode scale as $\sim n^{2j+1} $ ($\sim n^{2j+3} $) for $n>5$. Besides, we have observed that the scaling of $\Qjt(n)$ for $n<5$ is weaker than that of high refractive indices. We have also studied the effect of losses on the $Q$ factor of a mode. We have found that a single numerical parameter $\Bjt$ is needed to determine the $Q$ factor for any value of the extinction coefficient, i.e. $ \Qjt(n,k)=\dfrac{Q_{j,0}^{\tau}}{\vert   1 + B_j^{\tau}  k \cdot n^{2(j+1/2)+\tau} \vert}$ with  $Q_{j,0}^{\tau}$ being the $Q$ factor for the lossless case. We have shown that $\Bjt$ is almost constant for $n>5$, and have given its value for different modes. Finally, we have unveiled that $\Bjt$ is unequivocally linked to a negative value of the extinction coefficient (gain) which yields a infinite $Q$ factor. That is, the gain threshold that defines the lasing condition of a mode completely determines how the $Q$ factor of this mode is influenced by absorption losses. 

\textit{Acknowledgments.} This work was supported by a research grant (VIL50376) from VILLUM FONDEN.


%


\end{document}